# Freezing and quantization of current passing through a doubly connected superconductor with a point contact

V. P. Koverya, S. I. Bondarenko,  A. V. Krevsun, N. M. Levchenko, and I. S. Bondarenko

*B. I. Verkin Institute for Low-Temperature Physics and Engineering of the National Academy of Sciences of Ukraine, pr. Lenina 47, Kharkov 61103, Ukraine*

The particulars of dc current passage through a structure consisting of a doubly connected superconductor (DCS) with branches that are asymmetric with respect to length and critical current have been investigated experimentally. The short branch, which has the lowest critical current, was a clamping niobium-niobium point contact with length comparable to the coherence length of the superconductor. In contrast to a previously studied DCS with a short branch much longer than the coherence length, it was found that when the short-branch current reaches the critical value the currents in the branches of the DCS do not undergo self-excited oscillations; a current exceeding the critical value enters the long branch when this current is increased in portions (is quantized), and when it is subsequently decreased it freezes partially or completely in the DCS circuit.

PACS numbers: 74.25.F, 74.81.Fa

## I. INTRODUCTION

The physics of doubly connected superconducting structures is an important branch of fundamental and applied superconductivity. It is sufficient to mention the processes occurring in superconducting rings in a magnetic field[1] and in superconducting quantum interferometers.[2] Recently, we discovered that in a doubly connected superconductor (DCS), through which a constant transport current $I$ is passed, self-excited oscillations of the current arise when the critical current in one branch reaches a critical value. The DCS consisted of two branches with different critical currents and inductances (with inductance ratio 1:500). The length of the short branch, which had the lower critical current, was about 1 mm, and the inductance as $19 \cdot 10^{-8}$ H.

It is of interest to investigate how a decrease in the length of the short branch to a value comparable to the coherence length of superconductors used can affect the distribution of the current $I$ in the DCS when the critical current of this branch is reached.

The required branch can be made in the form of wither a cross-shaped film Josephson tunnel contact or a clamping superconducting point contact (SPC).[6] Because the fabrication technology is simpler we used SPC in our experiments.

The objective of the present work is to determine the distribution of the current $I$ in the branches of a DCS, when the current is introduced into it through a SPC, reaches and then exceeds its critical current.

## II. EXPERIMENTAL ARRANGEMENT

The electric circuit of the experimental DCS and the current-voltage characteristic (IVC) of the SPC used are displayed in Fig. 1.

The circuit of the doubly connected superconductor is made of 70 $\mu$m in diameter niobium microwire, whose ends were place on one another in a cross-shaped manner and clamped together mechanically by a metal clamp. An SPC arose at the point of intersection. The dc transport current 1 was fed from the free end of the niobium wire, as shown in Fig. 1a. The branch with the higher inductance $L_1$ $=5 \cdot 10^{-6}$ H was made in the form of coil with diameter 8 mm and $W=5$ loops. A detector of the magnetic field of the current flowing along the coil is placed inside the coil. The detector was a ferroprobe (FP) with sensitivity $10^{-5}$ Oe. The current of the coil was determined from the field, measured by the detector, using a pre-determined relation between the current in the coil and the indications of the FP. The inductance $L_2$ of the part of the DCS circuit with the SPC was assumed to be equal to the inductance of the contact. It was evaluated from the relation for the inductance of a Josephson contact $L_2=\Phi_0/\pi I_{c2}$ (Ref. 7) and was found to be $10^{-14}$ H for the contact studied in the present work. Thus the ratio of the inductances of the DCS branches was about $1:10^9$. The current $I$ could be set by a current source from $10^{-5}$ to 1 A. The main information on the current distribution in the branches of the DCS was obtained by detecting the magnetic field generated by the current flowing through the above-mentioned coil as the current $I$ increased and decreased. The DCS was immersed in liquid helium at $T=4.2$ K. The critical current, determined by the critical current of the SPC, of the branch with the lowest inductance in different samples of the circuit ranged from 20 to 120 mA, which was determined by the purity of the surface of the microwires at their clamping location and by the clamping force. The critical current and the current-voltage characteristics of the SPC, connected into the DCS circuit, were determined in the following sequence. First, the stability of the IVC of similar Nb-Nb clamping contacts with respect to the temperature cycling 300 K–4.2 K–300 K without including the contact in the DCS circuit was determined. It was found that for contacts with the initial critical current in the range 20–120 mA it changes after the cycles by no more than 5%, while the IVC shows no hysteresis to currents 150–180 mA. The contact described in this article was connected into the circuit, cooled together with it to $T=4.2$ K; its critical current $I_{c2}$ (found to be 80 mA) was measured according to the dependence in Fig. 2 (see below), after which the circuit was heated to $T=300$ K, cut, and re-cooled to $T=4.2$ K in order to measure the IVC of the contact, shown in Fig. 1b. It was found that the critical current was conserved and the IVC, to within the limits of its measurement accuracy, showed no hysteresis in measurements of the current through the contact from zero to 160 mA and from 160 mA to 0. The critical current of the niobium microprobe $I_{c1}$ and correspondingly of the branch with high inductance $L_1$ was about 4 A. The cryostat with the experimental DCS was screened from the Earth's magnetic field and its fluctuations by means of a magnetic screen. The amplitude of the residual low-frequency fluctuations of the surrounding magnetic field did not exceed $10^{-4}$ Oe, which corresponded to a change of the current in the coil with the ferroprobe by 10 $\mu$A. The IVC of the SPC which was not connected into the circuit and the dependences of the current in the coil with the FP on the current $I$ were recorded using an N-309 electromechanical automatic plotter.

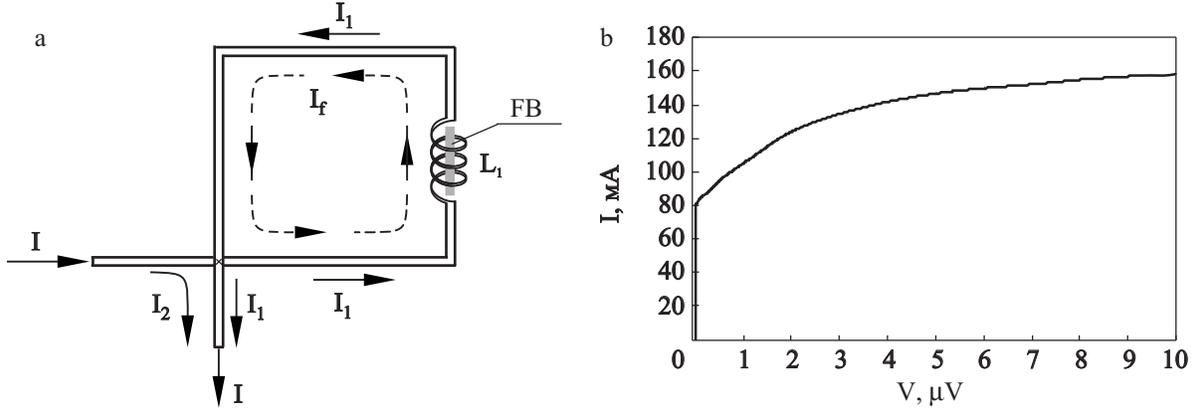

FIG. 1. a—Diagram of doubly connected superconductor (DCS) with a point contact, through which a constant transport current $I$ is passed; $I_1$ and $I_2$—currents in the branches of the superconducting circuit of DCS with inductances $L_1$ and $L_2$; $I_f$—frozen current in the DCS circuit; FB—ferroprobe for measuring the magnetic field of the current of the branch with conductance $L_1$; ×—position of the point contact between niobium conductors. b—Current-voltage characteristic of a point contact with an open DCS ring.

## III. EXPERIMENTAL RESULTS AND DISCUSSION

After the contact is connected into the circuit (Fig. 1a) and for transport current $I < 80$ mA the currents $I_1$ and $I_2$ in the branches of the DCS are distributed in accordance with the relation[6]

$$I_2/I_1 = L_1/L_2, \tag{1}$$

which for $L_1/L_2 \approx 10^9$ signifies that the transport current flows almost entirely through the SPC ($I_2 \approx I$). Starting at $I \cong I_{c2}$ a current $I_1$, growing in steps as $I$ increases from the point 1 to the point 7, appears in the branch 1, as shown in Fig. 2.

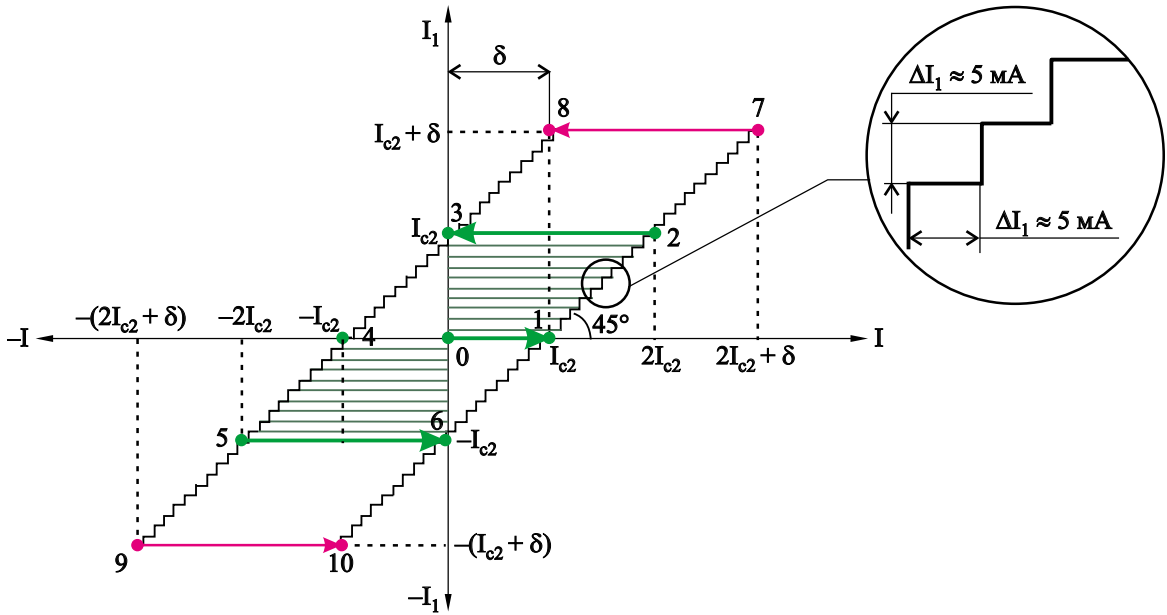

FIG. 2. Experimental dependence of the current $I_1$ in the branch No. 1 with the FB (branch with inductance $L_1$) versus the magnitude of the transport current $I$ through the DCS, $\delta$—some transport current added to $I=2I_{c2}$ such that the section 7–8 of the "plateau" is shifted relative to the axis $I_1$ by $\delta$, in contrast to similar "plateaus" arising as the current $I$ decreases from $I \lesssim 2I_{c2}$ (lines of several "plateaus" in the regions 1-2-3-0 and 4-5-6-0).

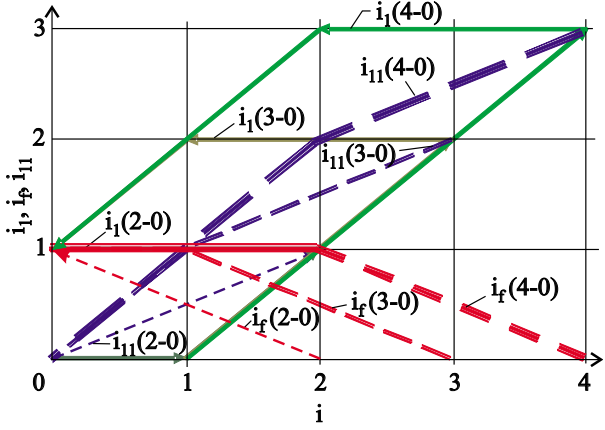

FIG. 3. Model dependences of the reduced (with respect to $I_{c2}$) values of the currents: total current in the branch No. 1 with the FB ($i_1 = I_1/I_{c2}$), parts of the current in the same branch in the form of a frozen current in the DCS circuit ($i_f = I_f/I_{c2}$) and part of the transport current of this branch ($i_{11} = I_{11}/I_{c2}$), decreasing in connection with a decrease of $I$, versus the magnitude of the reduced transport current $i = I/I_{c2}$. The figures in parentheses indicate the initial and final values of the decreasing transport current $i$. The arrows indicate the directions of change of the currents $i$ and $i_1$ with increasing current $i$ from zero to four and with it decreasing from $i = 2, 3, 4$, to zero.

In contrast to the DCS studied in Ref. 3 and 4, no self-excited oscillations of the current $I_1$ are observed when the critical current of the short branch of the DCS is reached ($I \geq I_{c2}$). The period ($\Delta I$) is several percent of $I_{c2}$, and $\Delta J \approx \Delta J_1$ (see inset in Fig. 2). The main part of the current $I_1$ increases (as compared with the size of the current steps) in the first quadrant of the coordinate system formed by the intersection of the axes $I$ and $I_1$ in proportion to the growth of the current $I$. As the current $I$, whose value is in the range $I = I_{c2} - 2I_{c2}$, decreases to zero the current $I_f = I_1$ freezes in the DCS circuit in the range $I_1 = 0 - I_{c2}$. We shall call this range of values of $I$ the first region. In the second region, for $I_{c1} + I_{c2} > I > 2I_{c2}$ (specifically, for $I = 2I_{c2} + \delta$, where $\delta$ is an increment to the current $I$ in Fig. 2), decreasing $I$ to zero likewise results in freezing of the current $I_1$, equal to $I_{c2}$. A feature of this region is the presence of a transitional section 8-3 with $I_1$ decreasing from the "plateau" with extent $2I_{c2}$ to current $I_1 = I_{c2}$ for $I = 0$. A change of the polarity and magnitude of the current $I$, starting at the frozen value of the current $I_f = I_{c2}$ at the point 3, as is evident in Fig. 2, make it possible to obtain the symmetric part of the function $I_1(I)$ passing through the points 4-5-9-10 in the third quadrant.

Let us first consider the processes associated with the change in the current of main, largest part of the current $I_1$ as a function of $I$, assuming that the stepped modulation of the current is substantially less than the critical current of the contact (as in the experiment). In this case sections with steps in the modulating process of change in $I_1$ can be replaced by lines (Fig. 3). Moreover, as will be shown below, for certain properties of the point contact the modulation $I_1(I)$ may absent completely. First, we shall consider the change of $I_1$ only as $I$ increases from $I = I_{c2}$. This section of the function $I_1(I)$ is a consequence of switching in the branch with inductance $L_1$ of the part of the current $I$ that exceeds $I_{c2}$. For this reason the angle between the axes of the function

$I_1(I)$ equals 45°. The linear growth of $I_1$ as $I$ increases is limited only by the critical current $I_{c1}$ of the branch with inductance $L_1$.

In the first quadrant of the function $I_1(I)$, i.e. for positive values of the currents $I$ and $I_1$, the behavior of the $I_1$ with decreasing $I$, starting with the appearance of $I_1$ and then from all of its large values, is of greatest interest.

The freezing of the current in the branch with $I$ decreasing to zero in the first region of its values can be explained by the appearance and conservation, after the current $I$ is switched off, of the undamped superconducting current $I_f$ in the DCS circuit (Fig. 1a), equal in magnitude to the transport current $I_1$ existing in the branch with FB before being switched off. The current $I_f$ is induced in the superconducting circuit closed through the SPC by virtue of the law of conservation of the magnetic flux created by the current $I_1$. In this circuit this arises by virtue of the fact that the indicated flux practically completely penetrates the area of the circuit, since the length of the circuit with the current $I_1$ is different from the length of the completely closed circuit only by the length of the SPC, comparable to the coherence length of the superconductor used. A similar freezing of the current in the circuit of the DCS, which has a short branch with size much greater than the large coherence length, did not occur[3-5] since before being switched off or decreasing the current passed only along the part of the DCS circuit and its magnetic flux would have appeared when the current flowed along the entire circuit.

As shown in Fig. 3, in the first region of values of $I$ the process described above explains the presence of a "plateau" in the function $I_1(I)$. For example, a decrease of the transport current from $i = 2$ to $i = 1$ and correspondingly the current in the branch No. 1 from $i_{11} = 1$ to $i_{11} = 0.5$ gives rise to the appearance of an equal induced undamped current $i_f = 0.5$, compensating the decrease of the current $i_{11}$, in the circuit. As a result the sum of the increments of these currents, which is recorded by the FP and equal to $i_1 = 1$ remains constant right up to $i = 0$. As one can see in Fig. 3, which shows that decrease of $i_{11}$ (2-0) and increase of the induced current $i_f$ (2-0) with $i$ varying from 2 to 0, the result of the indicated process is that the length of the "plateau" along the $i$ axis is 2, as is observed experimentally (Fig. 2). It is clear from the same considerations that when $i$ decreases from values less than 2 to 0 the magnitude of the plateau will between $i = 2$ and $i = 1$ and the current $i_f$ between 1 and 0. It should be noted that in this process the induced current $I_f$ flows in the SPC in a direction opposite to $I_2$, which does not branch in the circuit with the FB. Their algebraic sum is maintained at the level of the critical current $I_{c2}$. The current through the contact varies from $I_2 = I_{c2}$ at the start process resulting in the decrease of $I$ to $I_f = -I_{c2}$ at the end when $I = 0$.

The dependence $I_1(I)$ in the second region of the values of $I$ as it decreases is determined by a similar process, the only difference being that the maximum possible induced current, equal to the critical current of the contact, can no longer compensate the higher value of the current $I$ in the entire range where it decreases to zero. Thus, as when $I$ decreases, a "plateau" of the same length $2I_{c2}$ arises in the dependence $I_1(I)$, and its remaining section (8-3 in Fig. 2) represents the decrease of the current $I_1$ through the branch with the FB, equal to the decrease of the current $I$ to zero. In the dependence $i_1(i)$, shown in Fig. 3, this corresponds to a decrease of the current from $i_1=3$ to $i_1=1$ with $i$ decreasing from 2 to zero and $i_1=2$ to $i_1=1$ with $I$ decreasing from 1 to 0. The result is that for $I=0$ the current $I_f=I_{c2}$ obtains once again, which likewise corresponds to experiment.

The region of the "plateau" with a strictly stabilized value of the current $I_1$ in the branch No. 1 with the FB can be called the region of quasifrozen current, since the current $I_1$ in this branch is a sum of currents: undamped induced current $I_f$ and the transport current $I=I_{11}$. This sum is automatically maintained constant for any fluctuations of the current $I$ within the "plateau."

The function $I_1(I)$ in the region of the second, third, and fourth quadrants of Fig. 2 with a change of the direction of the current $I$ can be explained talking account of the conservation of the critical state of the SPC, appearance of induced current with different direction in the DCS circuit, freezing and quasifreezing of the current $I_1$ as considered for this dependence in the first quadrant. On this basis the reason for the absence of self-excited oscillations of the current in this type of DCS is understandable. It is explained by the fact that any decrease of the current $I_1$, which has entered the high inductance branch, as is typical for the process of self-excited oscillations in an different type of DCS,[3–5] is impossible here because of the quasifreezing of the current $I_1$ in the DCS circuit.

We shall now discuss the mechanism by which current steps appear in the dependence $I_1(I)$. The indicated modulation of this dependence can be explained using the microstructural features of the clamped SPC. As a rule, a real clamped superconducting contact between wires with natural surface nonuniformities is obtained at several points, as a result which the SPC is a superconducting quantum interferometer (SQI) with two or more point microcontacts of the Josephson type, connected in parallel, with different critical currents and separated by micron or submicron distances from one another.

The process occurring in a DCS with a short branch in the form of SQI can be explained on the basis of a simplified scheme of the DCS, shown in Fig. 4a. According to this scheme, it is supposed that the SQI has two identical point microcontacts 1, 2 along which a transport current $I=I_0$ in the range $I_0=0-I_{c2}$ can flow. When $I_0>I_{c2}$ a transport current greater than $I_{c2}$ can flow into the branch with the FB. The critical currents of each microcontact equal $I_{c0}$. The critical current $I_{c2}$ of a clamping contact in the form of SQI equals the sum of the critical currents of the microcontacts. The current $I$, which passes along the sections of the niobium wire, a part of the microcircuit of the SQI formed, creates a magnetic field $H$ which acts on the SQI. A change in $I$ results in a change of $H$ and the magnetic quantum modulation of the critical current SQI $I_{c2}(I)$ (Ref. 8) by the amount $\Delta J$. Thus in the present case a modulated part due to the dependence $I_{c2}(I)$ is superposed on the linear dependence $I_1(I)$ existing for $I>I_{c2}$ and typical for DCS without a SQI at the entrance. If $I<I_{c2}$, then the period $(\Delta H)$ of the quantum oscillations of $I_{c2}$ as a function of the field $H$ is

$$\Delta H = \Phi_0/\mu_0 S_0, \qquad (1')$$

where $\mu_0 = 4\pi \cdot 10^{-7}$ H, $S_0$ is the area of the circuit of quantization of SQI, and $\Phi_0$ is the quantum of magnetic flux. On the other hand the field produced by a transport current flowing along a conductor with the diameter $d$ on its surface can be estimated by means of a well-known relation describing the field of a long conductor with the current:

$$\Delta H \approx \Delta I/\pi d, \qquad (2)$$

where $\Delta I$ is the period of current oscillations. Since the current flows in the opposite a direction along two sections of the wire in the SQI circuit, the field increases by a factor of 2 as compared with the relation (2). As one can see in fig. 4a, the current generating the field in the SQI is only half of the transport current $I$ and for this reason the field must be decreased by a factor of 2. As a result the relation (2) correctly describes the field in the case $I < I_{c2}$. From the relations (1) and (2) we obtain

$$\Delta I \approx \pi \Phi_0 / \mu_0 (d/S_0). \qquad (3)$$

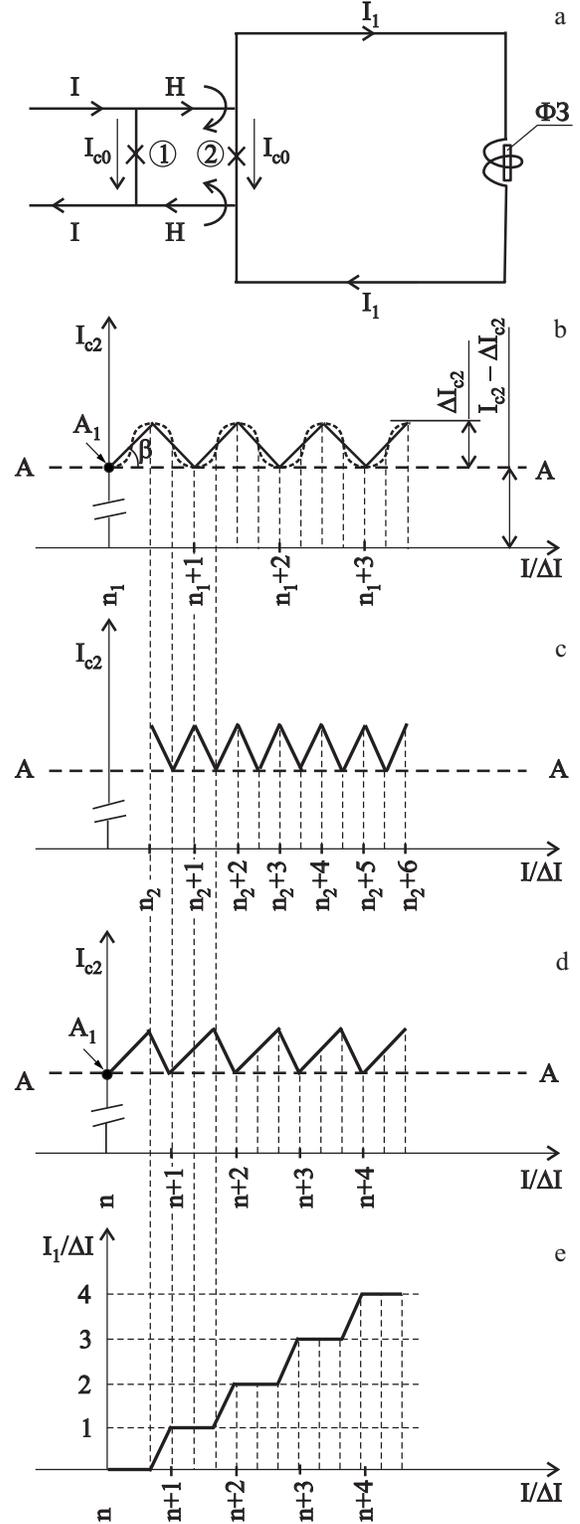

FIG. 4. a—Diagram of a doubly connected superconductor with transport current $I$ fed through a superconducting interferometer with two Josephson contacts (1, 2) with critical currents $I_{c0}$. b, c, d—Model dependences of the critical current $I_{c2}$ of an interferometer on the relative magnitude of the transport current $I/\Delta I$ for $I<I_{c2}$: $I \gtrless I_{c2}$ and with alternating values of the transport current from $I<I_{c2}$ to $I \gtrless I_{c2}$, respectively. e—Relative current in branch No. 1 with FB $(I_1/\Delta I)$ versus $I/\Delta I$, constructed on the basis of the dependence shown in Fig. 4d.

The modulation depth $\Delta I_{c2}$ can be estimated from the relation[2,8]

$$\Delta I_{c2} = \Phi_0 / 2L_0, \qquad (4)$$

where $L_0$ is the inductance of the SQI circuit. The form of the periodic function $I_{c2}(I)$ depends on different parameters of the SQI (specifically, on the current–phase characteristics of each contact and the critical currents of the contacts), which cannot be determined on the basis of the present work. For a qualitative explanation of the current steps arising in $I_1(I)$ (Fig. 4e) we shall confine ourselves to one possible form of $I_{c2}(I)$,[2,8] shown in Fig. 4b in the form of a rectified, for simplicity, construction of a sinusoid (shown by the dashed line), where several quanta of the change in $I_{c2}$ in the interval of quantum periods $I/\Delta I$ from $n_1$ to $n_1+3$, where $n_1$ is an integer. We shall examine the case of an increase of the current $I$ starting at the point $A_1$ in Fig. 4b, where $I$ is somewhat less than $I_{c2}$, i.e. $I = I_{c2} - \delta I$, where $\delta I \ll I_{c2}$. If the parameters of the SQI are such that $\tan \beta = \Delta I_{c2}/\Delta I \gtrsim 1$, then as $I/\Delta I$ increases from $n_1$ to $n_1+1/2$ the current will not flow into the branch with the FB, since it does not reach the critical magnitude and the SQI is in the superconducting state. In the dependence $I_1/\Delta I$ ($I/\Delta I$) a "plateau" will correspond to this increase of the current $I$ (Fig. 4e). When the current reaches the value $I/\Delta I = n_1 + 1/2$ the critical current of the SQI starts to decrease, the current $I$ through the SQI starts to exceed $I_{c2}$ ($I > I_{c2}$) and flows into the branch with the FB, we denote this current as $I_1$. In the process, the relation between the current flowing along both conductors and entering the SQI circuit, and field generated by it changes. Indeed, as one can see in Fig. 4a, the field $H$ is not produced by the total current $I$ and not by half the current, as was the case for $I < I_{c2}$. The field in the circuit can now be described by, instead of the relation (2), the relation

$$\Delta H \approx 2\Delta I/\pi d, \qquad (5)$$

and the period with respect to the current can be described by, instead the relation (3), the relation

$$\Delta I \approx \pi \Phi_0 / 2\mu_0 (d/S_0). \qquad (6)$$

Thus the dependence of the critical current of the SQI on the transport current passing through the branch with the FB has period which is half than that from the current passing only along the branch with the SQC. This is shown in Fig. 4c. As a result the quantum oscillations of the critical current of the SQI with a single period of the variation of $I_{c2}$ acquire an asymmetric form with a sharper (as compared with the oscillations in Fig. 4b) decrease of the critical current for $I/\Delta I > n_1 + 1/2$. In turn, this results in a sharp increase, in the form of a step, in the current $I_1$ in the dependence $I_1(I)$ (Fig. 4e). As the current increases further from the value $I/\Delta I = n_2 + 1/2$, because of the more rapid increase of the critical current of the SQI as compared with the increase of $I$ the interferometer once again becomes superconducting, and the current $I_1$ which previously arose in the DCS circuit, as follows from the explanation presented above for the form of the function $I_1(I)$, quasi-freezes, forming a "plateau" up to the next decrease of the critical current of the SQI. The oscillations of $I_{c2}$ being modeled are shown in a combined form in Fig. 4d. Evidently, Fig. 4e is similar to the experimentally observed stepped function $I_1(I)$ (Fig. 2). The smaller slope of the steps as compared with experiment could be due to the fact that the real point contacts in SQI have a current–phase characteristic that differs from a sinusoidal function, which we adopted in our model description of the processes in a DCS. Nonetheless, the experimentally measured values of the "plateau" and the height of the steps (see inset in Fig. 2) turned out to be close in magnitude, which corresponds to the condition for their appearance when $\Delta I_{c2}/\Delta I \gtrsim 1$ and the model dependence $I_1(I)$ in Fig. 4e.

Using the experimental values of $\Delta I$ and $\Delta I_{c2}$, the relations (3) and (4) can be used to estimate $S_0$ and $L_0$ of the interferometer, arising at the location of the SQI. They turned out to be equal to $10^{-13}$ m$^2$ and $2 \cdot 10^{-13}$ H, which confirms the supposition that the SQI is a submicron structure. It is also understandable on this basis that in the case when a single point of contact of microprobes in a clamping SQI an interferometer is not formed and steps with the indicated origin in $I_f(I)$ cannot arise, while the dependence itself will acquire a linear form which is identical to its model shown in Fig. 3.

## IV. CONCLUSION

The present investigations of the distribution of the transport current $I$ through an asymmetric doubly connected superconductor with a point contact in the role of a weak link of the DCS make it possible to determine the limit on the applicability of the principle of minimum magnetic energy for determining the current state of the DCS, demonstrate new methods of freezing current in superconducting closed circuits, and describe the physical processes occurring in a quantum structure in the form of a dc SQUID shunted by a superconducting inductance. We shall examine the main results indicated in greater detail.

As shown previously,[5] the time dependent distribution of the transport current in the branches of DCS when the critical current is reached in one of the macroscopic-size branches (i.e. with the size of a weak link, correspondingly exceeding the coherence length of the superconductors being used), having the form of self-excited oscillations (SEO) of the current is caused by the system in the form of the DCS striving to lower its magnetic energy to zero. However, if the transport current starts to enter the DCS through a branch in the form of a point contact (branch No. 2 in the present work), whose length is comparable to the coherence length ($\zeta_{Nb} \approx 40$ nm (Ref. 9)), then when the critical current $I_{c2}$ is reached the system responds to this completely differently. After part of the transport current flows into the branch No. 1 with a high critical current $I_{c1}$ its subsequent decrease in this branch and correspondingly a decrease of the magnetic energy of the system do not happen and SEO do not arise. Instead, the initial current system decomposes into two systems: the initial system in the form of a DCS with branching of the transport current along two branches and a second system without current branching, in the form of a closed circuit capable of carrying a frozen superconducting current, induced by any decreases of the current in the branch No. 1. A similar self-organizing current system is a quantum macroscopic system which possesses a set of allowed quantum current levels which correspond to the quantum of magnetic flux which is the only form that the flux created by the current can take in a superconducting closed circuit. Thus if the nonstationary distribution of the current in a DCS in which one of the macroscopic branches periodically passes into the critical state is a consequence of the classical striving of the system to reach the minimum magnetic energy, then in the case of a DCS in which there is one microscopic branch in the critical state, the current distribution in the branches is stationary and is determined by the allowed current quantum energy levels of a closed superconducting circuit formed by both branches of the DCS. As the transport current increases, the system passes into levels with increasingly higher magnetic energy. In the experimental DCS with macroscopic dimensions of the closed circuit these levels are split by very small energy gaps ($\delta\varepsilon = \Phi_0^2/2L_1 \approx 10^{-6}$ eV) and current gaps ($\delta I = \Phi_0/L_1 \approx 10^{-10}$ A) and for this reason current-step levels corresponding to them are not observed on the experimental curve $I_1(I)$.

As regards current freezing in DCS it is known that it possesses two advantages: current maintenance does not require energy input and it possesses record stability in time. Until now the following methods were known for freezing current in single—or multiloop superconducting rings and, in part, in multiply connected superconductors consisting of a HTSC ceramic:[10,11]

1) applying a magnetic field to a ring at temperature above the critical value, cooling the ring to a temperature below the critical value and switching the field off (FC—field cooling—method);
2) cooling a ring without an external field to a temperature below the critical value, applying to the ring a magnetic field above the critical value and switching the field off (ZFC—zero field cooling—method);
3) feeding the required transport current into an unclosed superconducting ring followed by closing the ring using a thermal superconducting switch and switching off the transport current.

The last method is most widely used in practice.

The present investigations of DCS with a point contact have made it possible to find another method of freezing a current in the range $0 - I_{c2}$ after a transport current, exceeding the critical current of the contact $I_{c2}$, is switched off as well as quasi-freezing a current in a superconducting circuit without switching off the transport current, realized in the range $0 - I_{c1}$, with stabilization of the quasi-frozen current within the range of variation of the transport current $\pm I_{c1}$. Neither of these methods of obtaining a stabilized current in a superconducting circuit requires the use of a special thermal switch; these methods permit continuous and even regulation of this current without feeding energy to the cryogenic agent.

Finally, the realization of a microscopic in length (comparable to the coherence length) branch of the DCS in the form of a superconducting interferometer has made it possible to determine that SQI, shunted by a superconducting inductance (inductance $L_1$ in the present experiment), plays the role of a quantum electronic valve which passes a transport current $I_1$, exceeding the critical current of the SQI, into the branch with inductance $L_1$ in portions (quanta). The parameters of the SQI determine the magnitude of the portions. Specifically, the height of the current steps in the function $I_1(I)$ is equal to the depth of magnetic modulation of the critical current of the SQI. It can be assumed that their slope is related with the form of the current–phase characteristic of the interferometer. To determine this, additional studies of DCS with an interferometer whose characteristics are known beforehand are required.